

\hsize     = 152mm
\vsize     = 215mm
\topskip   =  15pt
  \parskip   =   0pt
  \parindent =   0pt




\font\sixrm=cmr6
\font\sixi=cmmi6
\font\sixsy=cmsy6

\font\sevenrm=cmr7
\font\seveni=cmmi7
\font\sevensy=cmsy7

\font\twelverm=cmr12
\font\twelvei=cmmi12
\font\twelvesy=cmsy10 at 12pt
\font\twelveit=cmti12
\font\twelvesl=cmsl12
\font\twelvebf=cmbx12
\font\twelvett=cmtt12

\def\twelvepoint{%
\def\rm{\fam0\twelverm}%
\def\it{\fam\itfam\twelveit}%
\def\sl{\fam\slfam\twelvesl}%
\def\bf{\fam\bffam\twelvebf}%
\def\tt{\fam\ttfam\twelvett}%
\def\cal{\twelvesy}%
 \textfont0=\twelverm
  \scriptfont0=\sevenrm
  \scriptscriptfont0=\sixrm
 \textfont1=\twelvei
  \scriptfont1=\seveni
  \scriptscriptfont1=\sixi
 \textfont2=\twelvesy
  \scriptfont2=\sevensy
  \scriptscriptfont2=\sixsy
 \textfont3=\tenex
  \scriptfont3=\tenex
  \scriptscriptfont3=\tenex
 \textfont\itfam=\twelveit
 \textfont\slfam=\twelvesl
 \textfont\bffam=\twelvebf
 \textfont\ttfam=\twelvett
 \baselineskip=15pt
}

 \font\sixteenrm=cmr17

   \font\twentyrm=cmr10 scaled \magstep4


\hsize     = 152mm
\vsize     = 215mm
\topskip   =  15pt
  \parskip   =   0pt
\parindent =   0pt

\newskip\one
\one=15pt
\def\One{\vskip-\lastskip\vskip\one}

\newcount\LastMac
\def\Skipe{1}  
\def\Txe{2}    
\def\Hae{3}    
\def\Hbe{4}    

\def\SkipToFirstLine{
 \LastMac=\Skipe
 \dimen255=150pt
 \advance\dimen255 by -\pagetotal
 \vskip\dimen255
}

\def\Raggedright{%
 \rightskip=0pt plus \hsize
 \spaceskip=.3333em
 \xspaceskip=.5em
}

\def\Fullout{
 \rightskip=0pt
 \spaceskip=0pt
 \xspaceskip=0pt
}



\def\ct#1\par{
 \One
 \Raggedright
 \twentyrm\baselineskip=24pt
 #1
}

\def\ca#1\par{
 \One
 \Raggedright
 \sixteenrm\baselineskip=18pt
 \uppercase{#1}
}

\def\aa#1\par{
 \One
 \Raggedright
 \twelverm\baselineskip=15pt
 #1
}

\def\ha#1\par{
 \ifnum\LastMac=\Skipe \else \One\fi
 \LastMac=\Hae
 \Raggedright
 \twelvebf\baselineskip=15pt
 \uppercase{#1}
}

\def\hb#1\par{
 \LastMac=\Hbe
 \One
 \Raggedright
 \twelvebf\baselineskip=15pt
 #1
}

\def\tx{
 \ifnum\LastMac=\Hae \else
  \ifnum\LastMac=\Hbe \else
   \ifnum\LastMac=\Skipe \else \One
   \fi
  \fi
 \fi
 \LastMac=\Txe
 \Fullout
 \twelvepoint\rm
}

\def\bb{
 \One
 \Fullout
 \twelvepoint\rm
  \hangindent=1pc
  \hangafter=1
}


%


\let\miguu=\footnote
\def\footnote#1#2{{\parindent=10pt\baselineskip=13pt\miguu{#1}{#2}}}
\def\implies{\Rightarrow}
     \def\=>{\Rightarrow}
\def \==> {\Longrightarrow}




\def\ideq{\equiv}


\def\lto { {\raise1pt\hbox{$<$}} \!\!\!\! {\lower4pt\hbox{$\sim$}} }
\def\gto { {\raise1pt\hbox{$>$}} \!\!\!\! {\lower4pt\hbox{$\sim$}} }




\let\cc=\catcode
{\cc`\^^M=\active %
\gdef\losenolines{\cc`\^^M=\active \def^^M{\leavevmode\endgraf}}}
\def\literal {\begingroup \cc`\\=12 \cc`\{=12 \cc`\}=12 \cc`\$=12 \cc`\&=12
 \cc`\#=12 \cc`\%=12 \cc`\~=12 \cc`\_=12 \cc`\^=12 \cc`\*=12 \cc`\@=0
 \cc`\`=\active \losenolines \obeyspaces \tt}%
{\obeyspaces\gdef {\hglue.5em\relax}}

{\cc`\`=\active \gdef`{\relax\lq}}

\def\vquotingon{\cc`\"=\active}
\def\vquotingoff{\cc`\"=12}
\vquotingon
\def"{\literal\leavevmode\hbox\bgroup\com}

{\cc`\@=0 \cc`\\=12 @cc`@^^M=@active %
 @gdef@com#1"{#1@egroup@endgroup} %
 @gdef@thatisit^^M#1\endliteral{#1@endgroup@smallskip}}
\vquotingoff

\def\stacksymbols #1#2#3#4{\def\theguybelow{#2}
    \def\verticalposition{\lower#3pt}
    \def\spacingwithinsymbol{\baselineskip0pt\lineskip#4pt}
    \mathrel{\mathpalette\intermediary#1}}
\def\intermediary#1#2{\verticalposition\vbox{\spacingwithinsymbol
      \everycr={}\tabskip0pt
      \halign{$\mathsurround0pt#1\hfil##\hfil$\crcr#2\crcr
               \theguybelow\crcr}}}

\def\monthname {\ifcase\month\or January\or February\or March\or
April\or May\or June\or July\or August\or September\or October\or
November\or December\fi}

\newcount\mins  \newcount\hours  \hours=\time \mins=\time
\def\now{\divide\hours by60 \multiply\hours by60 \advance\mins by-\hours
     \divide\hours by60         
     \ifnum\hours>12 \advance\hours by-12
       \number\hours:\ifnum\mins<10 0\fi\number\mins\ P.M.\else
       \number\hours:\ifnum\mins<10 0\fi\number\mins\ A.M.\fi}


\def\<{\langle}
\def\>{\rangle}
\def\tr{{\rm tr}}

\ct
Impossible Measurements on Quantum Fields\footnote{{\it *}}\par
{{\noindent\parindent=10pt\baselineskip=12pt{{\it To appear in B. L. Hu and
T. A. Jacobson (eds.) ``Directions in General Relativity, Vol. II:
a Collection of Essays in honor of Dieter Brill's Sixtieth Birthday''
(CUP, 1993).}}}
}

\ca
Rafael D. Sorkin\par

\aa
Department of Physics, Syracuse University, Syracuse NY 13244-1130\par

\SkipToFirstLine

\tx
{\baselineskip=20pt
\centerline{\it Abstract}
{\baselineskip=13pt
{\leftskip=1.5cm\rightskip=1.5cm
It is shown that the attempt to extend the notion of ideal measurement
to quantum field theory leads to a conflict with locality, because
(for most observables) the state vector reduction associated with an
ideal measurement acts to transmit information faster than light.  Two
examples of such information-transfer are given, first in the quantum
mechanics of a pair of coupled subsystems, and then for the free
scalar field in flat spacetime.  It is argued that this problem leaves
the Hilbert space formulation of quantum field theory with no definite
measurement theory, removing whatever advantages it may have seemed to
possess vis a vis the sum-over-histories approach, and reinforcing the
view that a sum-over-histories framework is the most promising one for
quantum gravity.
\bigskip\bigskip}}}

\ha
1. Introduction: Ideal Measurements and Quantum Field Theory\par

\tx
Whatever may be its philosophical limitations, the textbook
interpretation of non-relativistic quantum mechanics is probably
adequate to provide the quantum formalism with all the predictive
power required for laboratory applications.  It is also
self-consistent in the sense that there exist idealized models of
measurements which allow the system-observer boundary to be displaced
arbitrarily far in the direction of the observer.  And the associated
``transformation theory'' possesses a certain formal beauty, seemingly
realizing the ``complementarity principle'' in terms of the unitary
equivalence of all orthonormal bases.  It is therefore natural to try
to generalize this semantic framework to relativistic quantum
field theory in the hope of learning something new, either from the
success or failure of the attempt.

\tx
In fact we will see that the attempt fails in a certain
sense; and the way it fails suggests that the familiar apparatus of
states and observables must give way to a more spacetime-oriented
framework, in which
the new physical symmetry implied by
the ``transformation theory'' is lost,
and the role of measurement as a fundamental concept is transformed or
eliminated.  Although such a renunciation of the usual measurement
formalism might seem a step backwards for quantum field theory, it can
also be viewed as a promising development for quantum gravity.  It
means that some of the conceptual problems which are normally thought
of as peculiar to quantum gravity or quantum cosmology are already
present in flat space, where their analysis and resolution may be
easier.

\tx
The framework we will attempt to generalize is based on the notion of
an {\it ideal measurement}, by which I mean one for which $(i)$ the
possible outcomes are the eigenvalues of the corresponding operator,
realized with probabilities given by the usual trace-rule, and $(ii)$
the standard {\it projection postulate} correctly describes the effect
of the measurement on the subsequent quantum state.  Such a
``minimally disturbing measurement'' is not only a ``detection'' but
is simultaneously a ``preparation'' as well and, if one follows the
textbook interpretation, it is precisely from this dual character of
measurement that the predictive power of the quantum mechanical
formalism derives.  It is also from this dual aspect that the
difficulty in relativistic generalization will arise---a difficulty
with so-called superluminal signaling.

\tx
Now in non-relativistic quantum mechanics, measurements are idealized
as occurring at a single moment of time.  Correspondingly the
interpretive rules for quantum field theory are often stated in terms
of ideal measurements which take place on Cauchy hypersurfaces.
However, in the interests of dealing with well-defined operators, one
usually thickens the hypersurface, and in fact the most general
formulations of quantum field theory assume that there corresponds to
any open region of spacetime an algebra of observables
which---presumably---can be measured by procedures occurring entirely
within that region.  (Unlike for the non-relativistic case, however,
no fully quantum models of such field measurements have been given, as
far as I am aware.)

\tx
The statement that ``one can measure'' a single observable $A$
associated to a spacetime region $O$ is fine as far as it goes, and an
obvious generalization of the projection postulate can be adopted as
part of the definition of such a measurement.  But a potential
confusion arises as soon as we think of two or more separate
measurements being made.  In the non-relativistic theory, measurements
carry a definite temporal order from which the logical sequence of the
associated state-vector reductions is derived; but in Minkowski space,
the temporal relationships among regions can be more complicated, and
the rules for ``collapsing'' the state
are not necessarily evident.  Nonetheless, I claim that a natural set
of rules exists, which directly generalizes the prescription of the
non-relativistic theory.

\ha 2. A Relativistic Projection Postulate\par

\tx
The problem is that these ``obvious'' rules fail to be consistent
with established ideas of causality/locality.  Hence, the kind of
measurements they envisage can presumably not be accomplished.  It
would of course be very interesting to try to construct models within
quantum field theory, to see what goes wrong, but I will not attempt
such a von-Neumann-like analysis here.  Also, I will restrict myself
to flat-space for definiteness, although nothing would be changed by
going over to an arbitrary globally hyperbolic spacetime.  Finally it
seems most convenient to work in the Heisenberg picture, since the
association of field operators to regions of spacetime is most direct
in that picture.

\tx
With these choices made, let us envisage a (finite) collection of
ideal measurements to be performed on some quantum field $\Phi$.  We
are then faced with a collection of regions, $O_k$ in Minkowski space,
and corresponding to each region, we are given an observable $A_k$,
formed from the restriction of $\Phi$ to $O_k$.  Given all this and an
initial state $\rho_0$ specified to the past of all of the $O_k$, we
may ask for the probability of obtaining any specified set of
eigenvalues $\alpha_k$ of the $A_k$ as measurement outcomes.

\tx
Non-relativistically, we would determine these probabilities by
ordering the $A_k$ in time (say with $A_1$ preceding $A_2$ preceding
$A_3 \, \ldots$), then
using $\rho_0$ to compute probabilities for the earliest observable
$A_1$, then ``reducing'' $\rho$ conditioned on the eigenvalue
$\alpha_1$, then using this reduced state to compute probabilities for
$A_2$, etc.  In the special case where each of the $A_k$ is a
projection or ``question'' $E_k$, this procedure results, as is well
known, in the remarkably simple expression
$$
  \<  E_1  E_2 \ldots E_{n-1} E_n E_{n-1}\ldots E_2 E_1  \>    \eqno(1)
$$
for the probability of `yes' answers to all the questions, where
expectation in the initial
state $\rho_0$ has been denoted simply by angle
brackets, a practice I will adhere to henceforth.  (That is, $\< A \>
\ideq {\rm trace}(\rho_0 \, A)$.)

\tx
Now let us return to the relativistic situation.  In the special case
where the regions $O_k$ are non-intersecting Cauchy surfaces (or
slight thickenings thereof), their time-ordering permits a unique
labeling, and the generalization of the above non-relativistic
procedure is immediate.  For a more general set of regions we may try
to foliate the spacetime in such a way that the $O_k$ {\it acquire} a
well-defined temporal ordering (each $O_k$ being separated from its
predecessor by one of the leaves of the foliation).  Obviously not
every labeling of the regions can arise from a foliation by Cauchy
surfaces, since no region which comes later with respect to such a
foliation can intersect the causal past of one which comes earlier.
Indeed this restriction makes sense independent of any choice of
foliation, and merely says that if a measurement made in one region
can possibly influence the outcome of a measurement made in a second
region, then the second measurement should be regarded as taking place
``later'' than the first.

\tx
The labelings of the regions which respect this causal restriction can
be described systematically in terms of an order relation ``$\prec$''
reflecting the possibilities of causal influence among the regions.
To define $\prec$ we merely specify that $O_j\prec O_k$ iff some point
of $O_j$ causally precedes some point of $O_k$.  A labeling --- or
equivalently a linear ordering --- of the regions is then {\it
compatible with} $\prec$ iff $O_j \prec O_k \implies j \leq k$.  Of
course, it can happen that no such labeling exists, in which case the
rules described above admit of no natural generalization.  To express
the exclusion of such cases in a systematic manner, we may take the
{\it transitive closure} of $\prec$, obtaining thereby an extended
relation, for which I will use the same symbol $\prec$.  The condition
that compatible labelings exist is then that this extended $\prec$ be
what is called a partial order, that is that it never happen that both
$O_j \prec O_k$ and $O_k \prec O_j$ for some $j \not= k$.  When, on
the contrary, this does happen, we have the analog of two
non-relativistic measurements being simultaneous, which, even
non-relativistically, leads to no well-defined probabilities unless
the corresponding operators happen to commute.  We may exclude such
cases, by simply requiring that the regions $O_k$ be disposed so that
no such circularity occurs.  In particular overlapping regions are
thereby excluded by fiat.

\tx
Given the partial order $\prec$, it is easy to state the natural
generalization of the non-relativistic rules for forming
probabilities: We simply extend $\prec$ to a linear order (as can
always be done) and use the rules precisely as they were stated
earlier.  Unless $\prec$ happens to be already a linear order, the
particular choice of linear extension (or equivalently labeling) is
not unique, but this ambiguity is harmless {\it as long as} the field
$\Phi$ satisfies ``local commutativity'', i.e.  as long as field
observables belonging to spacelike separated regions commute.  In
particular, when the $A_k$ are all projection-operators, the formula
(1) holds exactly as given above, for any labeling of the regions
which is compatible with $\prec$.

\tx
(It is often objected that the idea of state-vector reduction cannot
be Lorentz-in\-vari\-ant, since ``collapse'' will occur along
different hypersurfaces in different rest-frames.  However we have
just seen that well-defined probability rules can be given without
associating the successive collapses to any particular hypersurface.
Thus the objection is unfounded to the extent that one regards the
projection postulate as nothing more than a rule for computing
probabilities.  Of course if one takes the state-vector (or density
operator?) itself to be physically real, then the puzzle about
``where'' it collapses might remain.)

\tx
Abstractly considered, the scheme we have just taken the trouble to
construct seems impeccable, but in fact it has a problem foreshadowed
by our need to take a transitive closure in defining $\prec$.  The
problem is that the state-vector reduction implied by an ideal
measurement is non-local in such a way as to transmit observable
effects faster than light, something like an EPR experiment gone
haywire.  If we want to reject such superluminal effects, then we will
be forced to exclude the possibility of ideal measurements of most of
the spacetime observables we have been contemplating.

\ha 3. The Contradiction with Locality\par

\tx
In speaking of superluminal effects, the situation I have in mind
concerns three regions $O_1$, $O_2$ and $O_3$, situated so that some
points of the first precede points of the second, and some points of
the second precede points of the third, but all points of the first
and third are spacelike separated.  In fact, let us specialize $O_2$
to be a thickened spacelike hyperplane, with $O_1$ and $O_3$ being
bounded regions to its past and future respectively.  The
corresponding observables $A_1$, $A_2$, $A_3$, I will call `$A$',
`$B$' and `$C$' in order to save writing of subscripts.  The
non-locality (or ``acausality'') in question then shows up in the fact
that, for generic choices of the observables $A$, $B$, $C$ and of
initial state $\rho_0$, {\it the results obtained by measuring $C$
depend on whether or not $A$ was measured}, even though $A$ is
spacelike to $C$.  By arranging beforehand that $B$ will certainly be
measured, someone at $O_1$ could clearly use this dependence of $C$ on
$A$ to transmit information ``superluminally'' to a friend at
$O_3$.\footnote{$^1$}
{This effect is reminiscent of the acausality of [1], but
in that case, information is transmitted directly into the past,
rather than over spacelike separations, as here.  Also, the acausality
there depends on the assumption that one can directly ``observe''
spacetime properties of a history which need not correspond to any
traditionally defined operator in Hilbert space.  Here, in contrast,
even the traditionally defined observables give trouble.}

\hb A Simple Example with Coupled Systems \par

\tx
To see this effect at its simplest it may help to retreat from quantum
field theory to a more elementary situation, namely a pair of coupled
quantum systems together with three observables: $A$ belonging to the
first subsystem, $C$ belonging to the second, and $B$ being a joint
observable of the combined system.  As in the field-theory case, the
effect of measuring $B$ will be to make a prior intervention on the
first subsystem felt by the second.

\tx
To simplify the analysis further we can make a change which is
actually a generalization.  Instead of considering specifically a
measurement at $O_1$ [respectively, a measurement on the first
subsystem] we consider an arbitrary intervention, implemented
mathematically by a unitary operator $U$ formed (like $A$ itself) from
the restriction of $\Phi$ to $O_1$ [respectively a unitary element of
the observable algebra of the first subsystem].  That a measurement is
effectively a special case of this kind of intervention follows from
the observation that the effect of a measurement of the observable $A$
on the density-operator $\rho$ is to convert it into the
$\lambda$-average of $\exp(-i\lambda A) \rho \exp(i\lambda A)$, as a
one-line calculation will confirm.  (Here, of course, I mean the
effect of the measurement before the value of the result is taken into
account.)  Thus, measuring $A$ is equivalent to applying
$U=\exp(-i\lambda A)$ with a random value of the parameter $\lambda$,
and in this sense is a special kind of ``unitary intervention''.  It
follows in particular that if unitary intervention cannot transmit
information, then neither can any measurement.

\tx
Incidentally, the subsuming of measurement under unitary intervention
in this way leads to a more unified criterion for a theory to be
``local'': it is local if interventions confined to some region can
affect only the future of that region.  But it is interesting that,
although it no longer mentions measurement directly, this criterion is
still expressed in terms of intervention by an external agent.  As far
as I know there is no way to directly express the idea of locality in
the context of a completely self-contained system.

\tx
A specific example of the non-local influence in question is now easy
to obtain.  Let the two subsystems be spin-1/2 objects, and let their
initial state be $|dd\>$, where the two spin-states are $u$ and $d$.
At time $t_1$ let us ``kick'' the first subsystem (or more politely
``intervene'') by exchanging $u$ with $d$ (we apply the unitary
operator $\sigma_1$).  At time $t_2$ we measure the operator $B$ of
orthogonal projection onto the state $( |uu\> + |dd\>)/ \sqrt{2}$; and
at $t_3$ we measure an arbitrary observable $C$ of the second
subsystem.

\tx
It is straightforward to work out the density-operator which governs
this final measurement of $C$ by ``tracing out'' the first subsystem
from the state resulting from the $B$-measurement.  The result is that
the second subsystem ends up in the pure spin-down state, $|d\>\<d|$,
and the expectation-value of $C$ is accordingly $\<d|C|d\>$.  In
contrast, without the kick, we would have obtained an entirely
different effective state, namely the totally random density operator
$(1/2) (|u\>\<u| + |d\>\<d|)$, corresponding to a $C$-expectation of
$(1/2) \tr C$.  Thus, the detection of spin-up for the second
subsystem would be an unambiguous signal that the first subsystem had
not been kicked.

\tx
If you feel uneasy about using a kick rather than an actual
measurement, you can replace the intervention at $O_1$ with a
measurement of $A=\sigma_1$ (we apply $\exp(i\sigma_1\lambda)$ with a
random value of $\lambda$ instead of with $\lambda=\pi/2$).  The
computation is not much longer, and yields for the effective state of
subsystem 2, the density-operator $(1/4)|u\>\<u| + (3/4)|d\>\<d|$.
This again differs from the totally random state, though not so
strikingly as before.

\hb An Example in Quantum Field Theory \par

\tx
In a sense, the two subsystem example just given is all we need, since
one would expect to be able to embed it in any quantum field theory
which is sufficiently general to be realistic.  Still, one might worry
about non-localities having snuck in in connection with the particle
concept on which the identification of the subsystems would probably
be based in such an embedding; so it seems best to present an example
couched directly in terms of a quantum field and its observables.

\tx
The example will follow the lines of that just given, but the
computation is a bit more involved, and I will present it in slightly
more detail, working for convenience in the interaction picture, for
which the field $\phi$ evolves independently of the intervention, while
the state $\rho$ get ``kicked'' from its initial value $\rho_0$ to
$U \rho_0 U^*$, $U$ being the unitary operator which implements the kick.
  The quantum field will be a free scalar field $\phi(x)$ initially in
its vacuum state, and the three spacetime regions will be those
introduced above.  The kicking operator will be taken to be
$U=\exp(i\lambda\phi(y))$, where $y\in O_1$; and the observable
measured in $O_3$ will be $C=\phi(x)$, where $x\in O_3$. (Really the
fields should be smeared, but it will be clear that this would make no
difference.) Finally, the observable $B$ measured in $O_2$ (which can
be chosen as any operator at all, since $O_2$ includes the whole spacetime
in its domain of dependence) will be orthogonal projection onto the
state-vector
$$
      |b\> = \alpha |0\> + \beta |1\>,     \eqno(2)
$$
where $|0\>$ is the vacuum and $|1\>$ is some convenient one-particle
state; thus $B = |b\>\<b|$.
\smallskip

\tx
Denoting vacuum expectation values simply by $\< \cdot \>$, we may
express the mean value predicted for  $C$  as
$$
   \< U^* B C B U \> +  \< U^* (1-B) C (1-B) U \>,  \eqno(3)
$$
whose two terms correspond respectively to the outcomes 1 and 0 for
the $B$-measure\-ment.  [To derive this expression, we may begin with
the state $\rho=U\rho_0 U^*$, as it is
after the kick but before the
measurement of the projection $B$.  The probability of the outcome
$B=1$ is then $\tr\rho B = \tr \rho_0 U^* B U = \< U^* B U \>$, and
that of the outcome $B=0$ is $\<U^*(1-B)U\>$.  In the former case, the
projection postulate yields for the consequent (normalized) state,
$$
 \sigma = {B \rho B \over \tr B \rho B}
        = {B \rho B \over  \tr  \rho B},
$$
and therefore for the consequent expectation value of $C$,
$$
  {\rm Exp}(C|B=1) = \tr \, \sigma C = {\tr \rho B C B \over \tr \rho B}.
$$
When weighted with the probability $\tr\rho B$ of actually obtaining
$B=1$, this expression becomes the contribution of the $B=1$ outcome
to the final expectation-value of $C$:
$$
  {\rm Exp}(C,B=1) = \tr \rho BCB = \< U^*BCBU\>.
$$
Finally, adding in the contribution of the $B=0$ outcome yields (3),
as desired.]

\tx
In order that the mean value (3) not depend on the magnitude of the
kick, it is necessary in particular that its derivative with respect
to $\lambda$ vanish at $\lambda=0$ (or in other words that an
infinitesimal kick have no effect).  Now, this derivative may easily
be computed, and turns out to be (twice) the imaginary part of
$$
  \< \phi(y) (C + 2 BCB - BC - CB) \>,
$$
which therefore must be purely real in order that locality be
respected.  However the first and last terms are separately real; the
former equals $\< \phi(y) \phi(x) \>$, which is real because
$x\natural y$ (a notation meaning that $x$ is spacelike to $y$) whence
$\phi(x)\natural\phi(y)$ (a notation meaning that $\phi(x)$ and
$\phi(y)$ commute), and the latter reduces to $|\alpha|^2$ times the
former when the definition of $B$ is used.  With the aid of the
notation, $\psi(x)\ideq \<0|\phi(x)|1\>$, the results of combining the
two remaining terms can be written as
$$
  2 (\alpha^*\beta)^2 \, \psi(x)\psi(y) +
    (2|\alpha|^2-1)|\beta|^2\,\psi(x)^*\psi(y).
  \eqno(4)
$$
Here the star denotes complex conjugation, and the fact that $\phi(x)$
changes the particle number by $\pm 1$ has been used in places to
eliminate some terms which would have been present had $\phi(x)$ not
been a free field.

\tx
To show that (4) need not be real, we can, for example, eliminate its
second term by taking $|\alpha|^2=|\beta|^2=1/2$.  What remains can
then be given any desired phase by an appropriate choice of the
relative phase of $\alpha$ and $\beta$, unless it happens that
$\psi(x)$ or $\psi(y)$ vanishes.  Avoiding this possibility in our
choice of $\psi$ (indeed it would be difficult not to avoid it, since
$\psi$ is purely positive frequency!), we arrive at the conclusion
announced earlier.  Notice, incidentally, that we could have arranged
$\psi(\cdot)$ to manufacture a problem even with $\alpha=0$, but the
superposition with the vacuum state allows us to control the phase of
(4) for an arbitrary choice of the one-particle wave-function $\psi$.
On the other hand, setting $\alpha=0$ in (2) does have the advantage
of lending a particularly simple physical meaning to the measurement
$B$: it merely asks whether or not there is precisely one particle
present and if so whether that particle is in the specific state,
$|1\>$.
\vfill\break

\ha 4. Possible Implications for Quantum Field Theory and Quantum
       Gravity\par

\tx
In a way it is no surprise that a measurement such as of $B$, which
occupies an entire hypersurface, should entail a physical
non-locality; but surprising or not, the implications seem far from
trivial.  Unless one admits the possibility of superluminal signaling,
the entire interpretive framework constructed above for the quantum
field formalism must be rejected as it stands.  What then remains of
the apparatus of states and observables, on which the interpretation
of quantum mechanics is traditionally based?

\tx
A possible way to salvage our framework would be to further restrict
the allowed measurement-regions $O_j$  in such a manner that the
transitive closure we took in defining $\prec$ would be redundant.
For example, we could require that for each pair of regions $O_j$,
$O_k$, all pairs of points $x\in O_j$ and $y\in O_k$ be related in the
same way (i.e. either $x\natural y$ in all cases, or $x<y$ in all
cases, or $x>y$).  Such a restriction would block the kind of example
just presented, but it would also be a very severe limitation on
the allowed measurements (excluding measurements on Cauchy surfaces,
for example.)  Also it is difficult to see how the ability to perform
a measurement in a given region---or the effect of that measurement on
future probabilities---could be sensitive to whether some other
measurement was located totally to its past, or only partly to its
past and partly spacelike to it.\footnote{$^2$}
{On the other hand, there exists a purely formal consideration which
suggests that in fact there might be some difference between the two
cases.  If one demands an answer to ``where does the collapse
occur?'', the only viable response would seem to be ``along the past
light cone'', and that would indeed appear as an influence of a later
measurement on an earlier one, when the latter is partly spacelike to
the former.\smallskip }

\tx
Another way out might be to select the allowed measurements on some
more ad hoc basis than that which was set up in Section 2. For example
it can be shown, in the situation with the coupled subsystems, that
information transfer never occurs when $B$ is a {\it sum} of
observables, one belonging to each subsystem.  This suggests that one
might allow Cauchy-surface observables which were integrals of local
operators, even though other Cauchy-surface observables would still
have to be excluded.  Spatially smeared fields have this additive
character, for example (though they might be very singular as
operators), but fields which are also smeared in time do not.
Similarly, the most obvious gauge-invariant hypersurface observables
in nonabelian gauge theories like QCD, i.e. holonomies, are functions
of the commuting set of local variables $A_j(x)$, but they are not
linear functions.  This puts them in jeopardy because, in the two
subsystem situation, nonlinear combinations of pairs of observables do
in general lead to information transfer.  (For example, even something
as simple as the product of two projections has this
difficulty).\footnote{$^3$}
{An interesting problem would be to characterize which joint
observables of a pair of subsystems potentially lead to ``information
transfer'', and which don't. \smallskip}
This also portends trouble for diffeomorphism-invariant hypersurface
observables in quantum gravity.

\tx
This may be an appropriate place to comment on one of the few attempts I
know of to design concrete models of field measurements, namely that
of Bohr and Rosenfeld~[2].  Their idealized apparatus is designed to
measure averaged field values in an arbitrary pair of spacetime
regions (even overlapping ones!).  When their regions are related as
we required in Section 2, their results are consistent with the
conclusion that their procedure does indeed furnish ideal
measurements, as we defined this term earlier.  Specifically, the
apparatus interacts with the field only in the two specified regions,
and the uncontrollable disturbance exerted by the earlier measurement
on the later one is no greater than required by the commutator of the
corresponding operators.  It would thus seem important to extend (or
reinterpret) their essentially classical treatment of the apparatus to
a quantum one, in order to learn how close they come to actually
fulfilling the requirements for an ideal measurement.  Specifically,
one can ask whether they actually measure the field averages they
claim to, and whether the probabilities of the different possible
outcomes are those predicted by the quantum formalism (with special
reference to the use of the projection postulate after the first
measurement, since its effect could {\it only} be seen in a full
quantum treatment).  Such an analysis would be especially interesting
(even though Bohr and Rosenfeld only treat a free field) because there
is no obvious formal reason why their temporally smeared observables
should not suffer from the type of non-locality we have been
discussing here.

\tx
However such an analysis would turn out, though (and it doesn't look
that hard to do), there remains the more general fact that
the need to resort to a case-by-case analysis would still leave us
without
 any clear formal criterion for which
``observables'' can be ideally-measured, and which cannot; and we
might also be left without any general rule to take the place of the
projection postulate.  Moreover the charm of the ``transformation
theory'' would be lost as well, since different orthonormal bases would
no longer be equal before the Law of Locality (in the case of
hypersurface measurements, for example).

\tx
Now, as we are all aware, the question of what is the best dynamical
framework for quantum gravity is not one which everyone will answer in
the same way.  It is very possible that some as yet unknown framework
will be needed, but among existing interpretations of quantum
mechanics, I have long felt that the sum-over-histories is the most
promising, both for philosophical reasons and for practical ones.  It
has the great advantage that it deals with spacetime rather than just
space, so that what is usually called the ``problem of time'' hardly
makes an appearance.  In particular, notions like horizon-area are
well-defined, and quantum cosmology can investigate the early history
of the universe because the universe really does have a
history.\footnote{$^4$}
{Another approach which deserves mention here is that of David
Finkelstein, who views dynamics in terms of networks of elementary
processes of input/output or creation/annihilation, and
correspondingly draws a fundamental distinction between preparations
and detections.  Clearly this is relevant to the examples of Section
2, since our reasoning there required an ideal measurement to be both
a detection and a preparation at once. \smallskip}

\tx
In the non-relativistic context, however, the sum-over-histories has
the disadvantage of allowing you to ask ``too many'' questions,
including ones whose answering seems to lead to causality violations
similar to those of Section 3~[1].  And in the face of such difficulties,
one lacks a well-defined criterion to know in advance what
measurements are possible, or even what interactions should count as a
measurement.  In these ways, however, the sum-over-histories would now
appear to be no worse off than what survives of the
state-and-observable framework when one tries to extend it to flat
space quantum field theory (not to mention quantum
gravity!)\footnote{$^5$}
{The sum-over-histories also suffers esthetically from its inability
to incorporate the unitary symmetry of the so called transformation
theory.  But we have seen that the Hilbert-space framework is now no
better off in this respect either.}

\tx
With the formal notion of measurement compromised as it seems to be
already in quantum field theory, the
greatest advantage of the sum-over-histories may be that it does not
employ measurement as a basic concept.  Instead it operates with the
idea of a {\it partition} (or ``coarse-graining'') of the set of all
histories, and assigns probabilities directly to the members of a
given partition, using what I would call the quantum replacement for
the classical probability calculus.

\tx
Actually, there are (at least) two variants of this idea.  In the way
I like to think about it~[3], the partition is ``implemented'' by a
designated subsystem which gains information about the rest of the
universe.  In this way of thinking, the basic idea would be that of
probability {\it relative to} such a subsystem, and the difficulty is
that it is not fully
clear under what conditions information is actually
obtained.  (In practice, though, no problem is apparent in either of
the two extreme cases which the history of science has brought us so
far --- laboratory experiments and astronomical observations.)  In the
other approach~[4], the partition is given a priori, but subjected to
a condition of decoherence.  In this approach, the classical
probability calculus applies, and one effectively returns to a
classical (but stochastic) dynamical law.  Some drawbacks of this
variant are that one must do a very difficult analysis even to
authorize the use of probability, and (more fundamentally) that
decoherence is always {\it provisional}, since it applies to entire
histories and is therefore always in danger of being overturned by
future activities.

\tx
Let us hope that future activities by all of us will clarify some of
these issues, and bring out the new insights which further study of
the ``quantum measurement problem'' undoubtedly has to offer.  After
all, the dialectical inter-penetration of all existing things is at the
heart of the interpretation which Bohr wanted to give to the new
quantum formalism.  In the textbook formulation this inseparability
shows up in the impossibility of observing without also disturbing.
It will be interesting to see how it shows up in the modified
dynamical framework which will have to be developed for quantum
gravity, and, as we see now, perhaps even for quantum field theory.

\tx
In concluding I would like to dedicate this article to Dieter Brill,
in honor of his sixtieth birthday and in recollection of the many
happy hours I have spent discussing physics with him (and once in a
while playing sonatas together).  Happy Birthday, Dieter!


\tx
I would also like to thank John Friedman, Josh Goldberg, Jim Hartle
and David Malament for discussions and correspondence on the topic of
this paper.  This work was supported in part by NSF Grant No.
PHY-9005790.

\ha References\par
\bb
[1]
Sorkin, R.D., ``Problems with Causality in the Sum-over-histories
           Framework for Quantum Mechanics'', in A. Ashtekar and J. Stachel
           (eds.), {\it Conceptual Problems of Quantum Gravity} (Proceedings
           of the conference of the same name, held Osgood Hill, Mass., May
           1988), 217--227 (Boston, Birkh\"auser, 1991).
\bb
[2]
Bohr, N. and L. Rosenfeld,
``Zur Frage der Messbarkeit der Elektromagnetischen Feldgr\"ossen'',
{\it Det Kgl. Danske Videnskabernes Selskab., Mathematisk-fysiske
Meddelelser}, {\bf 12,} No. 8 (1933);
``Field and Charge Measurements in Quantum Electrodynamics'',
{\it Phys. Rev.} {\bf 78}:794-798 (1950).
\bb
[3]
 Sorkin, R.D., ``On the Role of Time in the Sum-over-histories
           Framework for Gravity'',
           paper presented to the conference on The History of Modern
           Gauge Theories, held Logan, Utah, July, 1987,
           to be published in
          {\it Int. J. Theor. Phys.} (1993, to appear);
 Sinha, Sukanya and R.D. Sorkin,
           ``A Sum-Over-Histories-Account of an EPR(B) Experiment''
            {\it Found. of Phys. Lett.}, {\bf 4}, 303-335, (1991).
\bb
[4]
For a review see:
Hartle, J.B., ``The Quantum Mechanics of Cosmology'', in {\it Quantum
Cosmology and Baby Universes: Proceedings of the 1989 Jerusalem Winter
School for Theoretical Physics}, eds. S. Coleman et al. (World
Scientific, Singapore, 1991)

\end